\documentclass[twoside,twocolumn,9pt]{article}
\usepackage{extsizes}
\usepackage[super,sort&compress,comma]{natbib}
\usepackage[version=3]{mhchem}
\usepackage[left=1.5cm, right=1.5cm, top=1.785cm, bottom=2.0cm]{geometry}
\usepackage{balance}
\usepackage{times,mathptmx}
\usepackage{sectsty}
\usepackage{graphicx}
\usepackage{lastpage}
\usepackage[format=plain,justification=justified,singlelinecheck=false,font={stretch=1.125,small,sf},labelfont=bf,labelsep=space]{caption}
\usepackage{float}
\usepackage{fancyhdr}
\usepackage{fnpos}
\usepackage[english]{babel}
\addto{\captionsenglish}{%
  
}
\usepackage{array}
\usepackage{droidsans}
\usepackage{charter}
\usepackage[T1]{fontenc}
\usepackage[usenames,dvipsnames]{xcolor}
\usepackage{setspace}
\usepackage[compact]{titlesec}
\usepackage{hyperref}
\usepackage{epstopdf}%This line makes .eps figures into .pdf - please comment out if not required.

\definecolor{cream}{RGB}{222,217,201}

\begin{document}

\pagestyle{fancy}
\fancypagestyle{plain}{

%%%HEADER%%%
%\fancyhead[C]{\includegraphics[width=18.5cm]{head_foot/header_bar}}
%\fancyhead[L]{\hspace{0cm}\vspace{1.5cm}\includegraphics[height=30pt]{head_foot/journal_name}}
%\fancyhead[R]{\hspace{0cm}\vspace{1.7cm}\includegraphics[height=55pt]{head_foot/RSC_LOGO_CMYK}}
\renewcommand{\headrulewidth}{0pt}
}
%%%END OF HEADER%%%

%%%PAGE SETUP - Please do not change any commands within this section%%%
\makeFNbottom
\makeatletter
\renewcommand\LARGE{\@setfontsize\LARGE{15pt}{17}}
\renewcommand\Large{\@setfontsize\Large{12pt}{14}}
\renewcommand\large{\@setfontsize\large{10pt}{12}}
\renewcommand\footnotesize{\@setfontsize\footnotesize{7pt}{10}}
\makeatother

\renewcommand{\thefootnote}{\fnsymbol{footnote}}
\renewcommand\footnoterule{\vspace*{1pt}%
\color{cream}\hrule width 3.5in height 0.4pt \color{black}\vspace*{5pt}}
\setcounter{secnumdepth}{5}

\makeatletter
\renewcommand\@biblabel[1]{#1}
\renewcommand\@makefntext[1]%
{\noindent\makebox[0pt][r]{\@thefnmark\,}#1}
\makeatother
\renewcommand{\figurename}{\small{Fig.}~}
\sectionfont{\sffamily\Large}
\subsectionfont{\normalsize}
\subsubsectionfont{\bf}
\setstretch{1.125} %In particular, please do not alter this line.
\setlength{\skip\footins}{0.8cm}
\setlength{\footnotesep}{0.25cm}
\setlength{\jot}{10pt}
\titlespacing*{\section}{0pt}{4pt}{4pt}
\titlespacing*{\subsection}{0pt}{15pt}{1pt}
%%%END OF PAGE SETUP%%%

%%%FOOTER%%%
\fancyfoot{}
%\fancyfoot[LO,RE]{\vspace{-7.1pt}\includegraphics[height=9pt]{head_foot/LF}}
%\fancyfoot[CO]{\vspace{-7.1pt}\hspace{13.2cm}\includegraphics{head_foot/RF}}
%\fancyfoot[CE]{\vspace{-7.2pt}\hspace{-14.2cm}\includegraphics{head_foot/RF}}
\fancyfoot[RO]{\footnotesize{\sffamily{1--\pageref{LastPage} ~\textbar  \hspace{2pt}\thepage}}}
\fancyfoot[LE]{\footnotesize{\sffamily{\thepage~\textbar\hspace{3.45cm} 1--\pageref{LastPage}}}}
\fancyhead{}
\renewcommand{\headrulewidth}{0pt}
\renewcommand{\footrulewidth}{0pt}
\setlength{\arrayrulewidth}{1pt}
\setlength{\columnsep}{6.5mm}
\setlength\bibsep{1pt}
%%%END OF FOOTER%%%

%%%FIGURE SETUP - please do not change any commands within this section%%%
\makeatletter
\newlength{\figrulesep}
\setlength{\figrulesep}{0.5\textfloatsep}

\newcommand{\topfigrule}{\vspace*{-1pt}%
\noindent{\color{cream}\rule[-\figrulesep]{\columnwidth}{1.5pt}} }

\newcommand{\botfigrule}{\vspace*{-2pt}%
\noindent{\color{cream}\rule[\figrulesep]{\columnwidth}{1.5pt}} }

\newcommand{\dblfigrule}{\vspace*{-1pt}%
\noindent{\color{cream}\rule[-\figrulesep]{\textwidth}{1.5pt}} }

\makeatother
%%%END OF FIGURE SETUP%%%

%%%TITLE, AUTHORS AND ABSTRACT%%%
\twocolumn[
  \begin{@twocolumnfalse}
\vspace{3cm}
\sffamily
\begin{tabular}{m{4.5cm} p{13.5cm} }

& \noindent\LARGE{\textbf{Pattern formation in two-dimensional hard-core/soft-shell systems with variable soft shell profiles}} \\%Article title goes here instead of the text "This is the title"
\vspace{0.3cm} & \vspace{0.3cm} \\

 & \noindent\large{Walter R.~C. Somerville,\textit{$^{a,b \ddag}$} Adam D. Law,\textit{$^{c \ddag}$} Marcel Rey,\textit{$^{d,e}$} Nicolas Vogel,\textit{$^{d,e}$} Andrew J. Archer,\textit{$^f$} and D. Martin A. Buzza$^{\ast}$\textit{$^{a,g}$}} \\%Author names go here instead of "Full name", etc.

& \noindent\normalsize{
Hard-core/soft shell (HCSS) particles have been shown to self-assemble into a remarkably rich variety of structures under compression due to the simple interplay between the hard-core and soft-shoulder length scales in their interactions. Most studies in this area model the soft shell interaction as a square shoulder potential. Although appealing from a theoretical point of view, the potential is physically unrealistic because there is no repulsive force in the soft shell regime, unlike in experimental HCSS systems. To make the model more realistic, here we consider HCSS particles with a range soft shell potential profiles beyond the standard square shoulder form and study the model using both minimum energy calculations and Monte Carlo simulations. We find that by tuning density and the soft shell profile, HCSS particles in the thin shell regime (i.e., shell to core ratio $r_1/r_0 \leq \sqrt{3}$) can form a large range of structures, including hexagons, chains, squares, rhomboids and two distinct zig-zag structures. Furthermore, by tuning the density and $r_1/r_0$, we find that HCSS particles with experimentally realistic linear ramp soft shoulder repulsions can form honeycombs and quasicrystals with 10-fold and 12-fold symmetry. Our study therefore suggests the exciting possibility of fabricating these exotic 2D structures experimentally through colloidal self-assembly.
} \\%The abstrast goes here instead of the text "The abstract should be..."

\end{tabular}

 \end{@twocolumnfalse} \vspace{0.6cm}

]

%%%END OF TITLE, AUTHORS AND ABSTRACT%%%

%%%FONT SETUP - please do not change any commands within this section
\renewcommand*\rmdefault{bch}\normalfont\upshape
\rmfamily
\section*{}
\vspace{-1cm}

%%%FOOTNOTES%%%

\footnotetext{\textit{$^{a}$~G. W. Gray Centre for Advanced Materials, Department of Physics \& Mathematics, University of Hull, Hull HU6 7RX, United Kingdom. }}
\footnotetext{\textit{$^{b}$~The MacDiarmid Institute for Advanced Materials and Nanotechnology, School of Chemical and Physical Sciences, Victoria University of Wellington, Kelburn Parade, Wellington 6012, New Zealand. }}
\footnotetext{\textit{$^{c}$~medPhoton GmbH, Strubergasse 16, 5020 Salzburg, Austria. }}
\footnotetext{\textit{$^{d}$~Institute of Particle Technology, Friedrich-Alexander University Erlangen-Nuernberg, Cauerstrasse 4, 91058 Erlangen, Germany. }}
\footnotetext{\textit{$^{e}$~Interdisciplinary Center for Functional Particle Systems, Friedrich-Alexander University Erlangen-Nuernberg, Ha-berstrasse 9a, 91058 Erlangen, Germany. }}
\footnotetext{\textit{$^{f}$~Department of Mathematical Sciences, Loughborough University, Loughborough LE11 3TU, United Kingdom. }}
\footnotetext{\textit{$^{g}$~E-mail: d.m.buzza@hull.ac.uk }}

%Please use \dag to cite the ESI in the main text of the article.
%If you article does not have ESI please remove the the \dag symbol from the title and the footnotetext below.
%\footnotetext{\dag~Electronic Supplementary Information (ESI) available: [details of any supplementary information available should be included here]. See DOI: 00.0000/00000000.}
%additional addresses can be cited as above using the lower-case letters, c, d, e... If all authors are from the same address, no letter is required

\footnotetext{\ddag~These authors contributed equally to this work}

%%%END OF FOOTNOTES%%%

%%%MAIN TEXT%%%%

\section{Introduction}
One of the distinguishing features of soft matter systems is the presence of a wide variety of interactions. The subtle interplay between the different types of interactions leads to a very rich self-assembly behaviour. An outstanding example of this is the case of hard-core/soft-shell (HCSS) particles which interact through an isotropic pair potential consisting of two characteristic length-scales, a short range hard-core potential and a longer-range soft shoulder repulsion (see Figure 1(a)). Experimentally, such a potential can arise for example in microgel particles with a higher cross-link density in the core compared to the corona\cite{Rey2016,Geisel2014}, rigid colloidal particles decorated with a soft polymeric layer\cite{Vogel2012,Rey2017,Rey2018,ElTawargy2018}, or block copolymer micelles consisting of a core of hydrophobic blocks surrounded by a shell of hydrophilic blocks.\cite{Fischer2011}

Jagla\cite{Jagla1998,Jagla1999} was the first to show that when HCSS particles are compressed in two dimensions (2D), they can self-assemble into a surprisingly rich variety of \emph{anisotropic} structures beyond the simple hexagonal structures that one would expect for spherical particles with isotropic interactions. These structures include stripes, squares, rhomboids and other complex crystal structures containing more than one particle per unit cell. Since the pioneering work of Jagla, a great variety of other structures have been found theoretically for HCSS particles in 2D including labyrinths,\cite{Camp2003} cluster and inverse cluster phases\cite{Malescio2003,Glaser2007,Fornleitner2008,Fornleitner2010} and quasicrystals of various symmetries.\cite{Jagla1998,Dotera2014,Pattabhiraman2015,Schoberth2016,Pattabhiraman2017} Remarkably, this rich variety of patterns is generated from a simple competition between the hard-core and soft-shoulder length scales in the interaction.\cite{Rey2017,Glaser2007,Dotera2014,Archer2013} Specifically, when the core-shell particles are compressed such that their shells begin to touch, provided the profile of the soft repulsive shoulder is flat enough so that the energy difference between fully and partially overlapping shells is small, the system can minimise its energy by fully overlapping neighbouring shells in some directions (i.e. so that the cores touch) in order to prevent the overlap of shells in others (i.e., so that the shells touch).

Most theoretical studies of HCSS particles in the literature model the soft shell interaction as a square shoulder potential. Although appealing from a theoretical point of view, this potential is physically unrealistic in the sense that it does not generate a repulsive force in the soft shell regime because the potential is flat.\cite{Schoberth2016} This situation is in contrast to real HCSS systems, e.g., colloids decorated by grafted polymers, where the interaction potential is not flat\cite{Witten1986,Milner1988,Norizoe2005}, and a finite repulsive force therefore exists when the soft shells of neighbouring particles overlap. To overcome this, here we consider HCSS particles with a range soft shell potential profiles beyond the standard square shoulder potential, including the linear ramp potential which is experimentally more realistic.\cite{Rey2017} Note that a recent study by Schoberth et al\cite{Schoberth2016} also considers HCSS particles with non-square soft shoulder interactions, although the range of soft shoulder potentials considered in that study are different from the ones considered here.

To simplify our discussion, we focus on HCSS particles in the `thin' shell regime, which we define as the case where the soft shoulder to hard core length scale ratio $r_1/r_0 < \sqrt{3}$. The significance of $\sqrt{3}$ in this context is that shell overlaps beyond nearest neighbour particles are only possible when $r_1/r_0 > \sqrt{3}$. We study the self-assembly of two-dimensional HCSS particles in this regime by performing both minimum energy calculations and finite temperature Monte Carlo (MC) simulations.

For our minimum energy calculations, our goal is to find a minimal model that captures the essential physics for the self-assembly of the system into periodic structures. In ref.\cite{Rey2017,Rey2018}, we performed a preliminary analysis where we considered minimum energy configurations (MECs) containing only one particle per unit cell. The one-particle model is able to reproduce the hexagonal, square and chain structures that are found experimentally in ref.\cite{Rey2017,Rey2018}, but it is not capable of generating the more complex periodic structures found in more elaborate models.\cite{Jagla1998,Jagla1999,Fornleitner2008,Fornleitner2010} In this paper, we therefore extend our calculations to include structures that contain up to two particles per unit cell.

Surprisingly, we find that this simple extension is sufficient to reproduce the periodic structures found from more sophisticated Genetic Algorithm calculations (which include up to 15 particles per unit cell).\cite{Fornleitner2008,Fornleitner2010} The MECs we find are also corroborated by our MC simulations. This good agreement demonstrates that our two-particle model is sufficient to capture many of the periodic structures that are accessible to HCSS particles in the thin shell regime. Our results for the phase diagram reproduce the broad features found by Jagla in his seminal studies,\cite{Jagla1998,Jagla1999} but correct a number of important discrepancies in those studies arising from the incomplete set of MECs that were used.\cite{Jagla1999} The model also allows us to predict the conditions under which HCSS particles in the thin shell regime can form the honeycomb lattice.

In addition to using MC simulations to check the accuracy of our minimum energy calculations, the MC simulations also allow us to study non-periodic structures such as quasicrystals. Specifically, by tuning both the ratio $r_1/r_0$ and density, we show that HCSS particles with a linear ramp soft shoulder potential can form dodecagonal (i.e., 12-fold symmetric) and decagonal (i.e., 10-fold symmetric) quasicrystals. These results are in excellent agreement with previous studies of HCSS particles with a square shoulder repulsion.\cite{Dotera2014,Pattabhiraman2015} However, since linear ramp potentials are more experimentally realistic,\cite{Rey2017} our study suggests the exciting possibility of forming these 2D quasicrystalline structures experimentally through colloidal self-assembly.

\section{Theoretical Methods}

\subsection{Hard-Core/Soft-Shoulder Potential}

We assume that the particles in our system interact through the generic hard-core/soft-shoulder potential proposed by Jagla\cite{Jagla1999}
\begin{equation}\label{hcss}
  U_g(r)=   \begin{cases}
                \infty, & r < r_0 \\
                U_0 \frac{g+\left[ \left( \frac{r-r_0}{r_1-r_0} \right) (g-g^{-1})-g \right]^{-1}}{g-g^{-1}}, & r_0 \leq r \leq r_1 \\
                0,  & r>r_1
             \end{cases}
\end{equation}
where $r_0$, $r_1$ are the range of the hard-core and soft-shoulder repulsion respectively, $U_0$ is the shoulder height and the parameter $g$ controls the profile of the soft shoulder, going from no shoulder ($g=0$), via a linear ramp ($g=1$) to a square shoulder ($g=\infty$) as we increase $g$ (see Figure 1(a)). As mentioned in the Introduction, we consider thin shell HCSS particles in this paper where $r_1/r_0 < \sqrt{3}$. Note that most theoretical studies of HCSS particles in the literature focus on hard-core/square shoulder potentials, i.e., $g=\infty$. In this paper, we consider variable $g$ in order to study the impact of $g$ on the phase behaviour of HCSS particles.

\subsection{Minimum Energy Calculations}

In order to determine the equilibrium structures formed by HCSS particles when they are compressed in two dimensions, we first calculate the minimum energy configurations (MECs) of the system, i.e., the equilibrium structure at zero temperature. The zero temperature regime is relevant so long as the energy scale of the soft-shell repulsion is much greater than the thermal energy, i.e., $U_0 \gg k_B T$, or equivalently the reduced temperature $T^* \equiv k_B T/U_0 \ll 1$, a condition that is easily satisfied in many experimental systems.\cite{Rey2017,Rey2018} As mentioned in the Introduction, we perform a comprehensive exploration of all two-dimensional structures containing (up to) two particles per unit cell. Specifically, we can define the unit cell as a parallelogram spanned by two lattice vectors $\mathbf{a}=a(1,0)$, $\mathbf{b}=a \gamma (\cos \phi, \sin \phi)$, where $\phi$ is the angle between the lattice vectors, $\gamma = b/a$ is the aspect ratio of the unit cell and $a,b$ are the lattice constants (see Figure 1(b)). Within this unit cell, the first particle is at $(0,0)$ (without loss of generality) while the second particle is at $\mathbf{r}=\alpha \mathbf{a} + \beta \mathbf{b}$, where $\alpha,\beta \in (0,1)$ are the coordinates of the second particle in the lattice basis set.

When calculating the zero temperature phase diagram, it is convenient to work in the NPT ensemble where the area per particle is variable. This is because the system exists as a single phase in the NPT ensemble except at the coexistence pressure between two or more phases. Specifically, parameterising the area per particle as $\frac{\sqrt{3}}{2} \ell^2$, where $\ell$ is the lattice constant of the system in the hexagonal phase, and noting that the area per unit cell is $a^2 \gamma \sin \phi$, for two particles per unit cell, the lattice constant $a$ is fixed by the condition $(a^2 \gamma \sin \phi)/2=\frac{\sqrt{3}}{2} \ell^2$. We can therefore express $a$ as a function of $\ell$, $\gamma$ and $\phi$ as $a=\ell \left(\frac{\sqrt{3}}{\gamma \sin \phi} \right)^{1/2}$. Note that the number density of HCSS particles (i.e., number of particles per unit area) is given in terms of the parameter $\ell$ by $\rho = 2/(\sqrt{3} \ell^2)$ while the core area fraction is given by $\eta=\pi r_0^2/(2 \sqrt{3} \ell^2)$.

\begin{figure*}
 \centering
 \includegraphics[width=1.8\columnwidth]{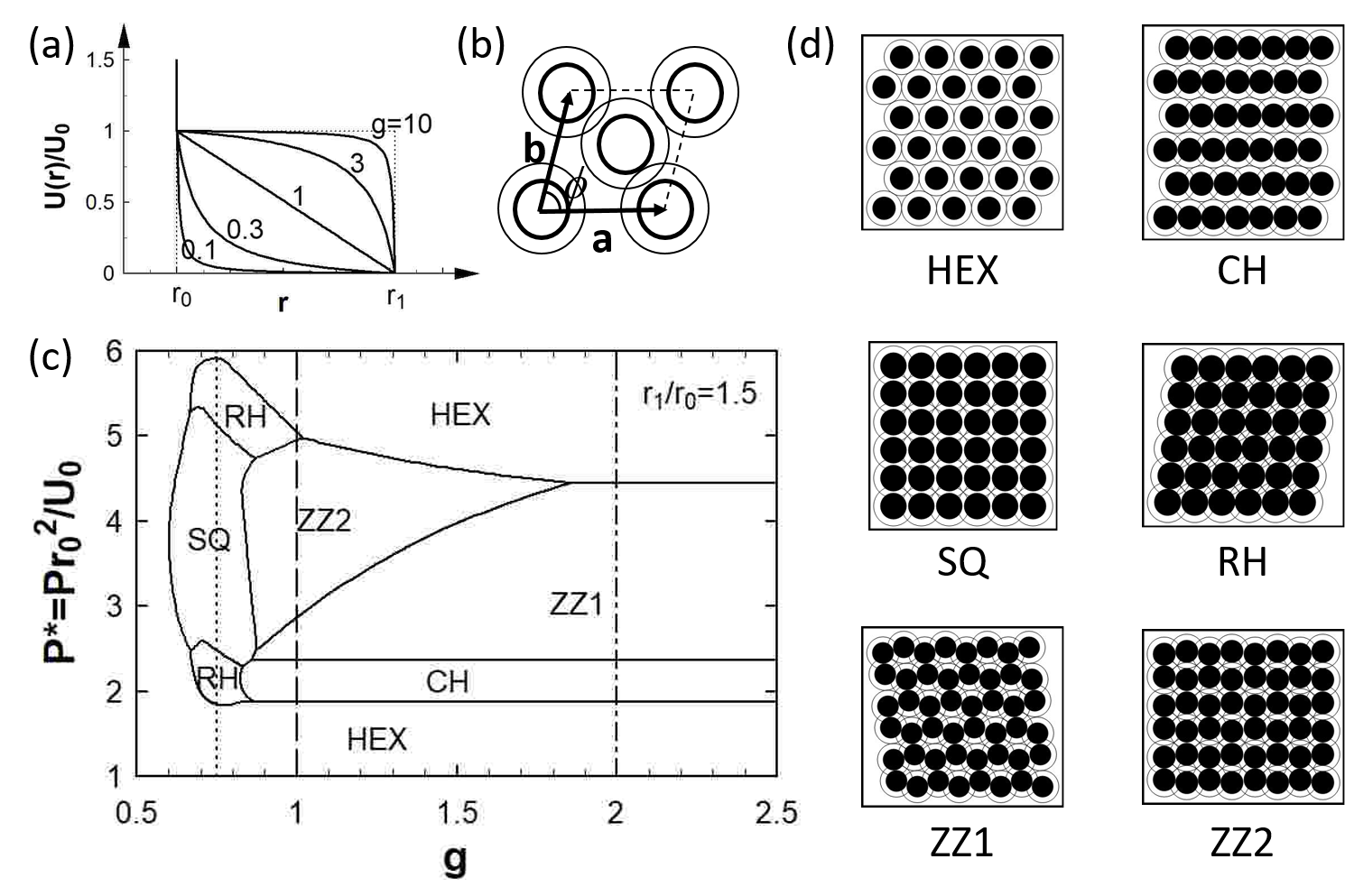}
 \caption{(a) Jagla potential with energy scale $U_0$, hard-core range $r_0$, soft shell range $r_1$ and different values for the soft shell profile parameter $g$. The dotted lines on the left and upper right correspond to $g=0$ (no shoulder) and $g=\infty$ (square shoulder), respectively. (b) Unit cell containing two particles used in our minimum energy calculations, where $\mathbf{a}, \mathbf{b}$ are the lattice vectors, $\phi$ is the unit cell angle and the thick and thin circles represent the particle core and corona, respectively. (c) Zero temperature phase diagram for HCSS particles with $r_1/r_0=1.5$ in the $g$ and reduced pressure $P^*$ plane. The dotted, dashed and dotted dashed vertical lines correspond to the state points for which we performed Monte Carlo simulations in Figure 2. (d) Representative minimum energy configurations (MECs) for $r_1/r_0=1.5$. Note that the hexagonal structure (HEX) shown is the low density hexagonal lattice (i.e., HEXL, no overlap of shells).}
 \label{figure1}
\end{figure*}

In the NPT ensemble, the equilibrium state is found by minimising the Gibbs free energy, or more specifically by minimising the enthalpy when we are at zero temperature. For our HCSS particles, the enthalpy per particle for crystal structures containing one particle per unit cell is given in ref.\cite{Rey2017}, while for crystals containing two particles per unit cell it is
\begin{equation}\label{gibbs2}
  H = \frac{1}{2} \left[ \sum_{h,k \neq 0} U_g (|h \mathbf{a} + k \mathbf{b}|) + \sum_{h,k} U_g (|h \mathbf{a} + k \mathbf{b}+\alpha \mathbf{a} + \beta \mathbf{b}|) \right] + \frac{\sqrt{3}}{2} P \ell^2
\end{equation}
where $U_g (r)$ is the HCSS potential given by Eq.~\eqref{hcss}. The first term on the right hand side of Eq.\eqref{gibbs2} represents the interaction energy per particle calculated via lattice sums, while the second term is the surface pressure $P$ (i.e., force per unit length) times the area per particle. Specifically, the first and second sum inside the square brackets represent the interactions between like particles (i.e., particle 1--particle 1 or particle 2--particle 2) and unlike particles (i.e., particle 1--particle 2) respectively and the factor $\frac{1}{2}$ outside the square bracket converts the interaction energy per unit cell to the interaction energy per particle. Both summations run over all integer values of $h,k$ satisfying $|h \mathbf{a}+k \mathbf{b}|\leq r_c$ (except for $h,k=0$ in the first sum), where $r_c$ is the cut-off radius for interactions. Since we are considering the thin shell regime where $r_1/r_0 < \sqrt{3}$, we are able to use a very short cut-off length of $r_c/r_0=2$.

In order to calculate the zero temperature phase diagram, we determined the MECs of the system as a function of $r_1/r_0$, $g$ and $P$ (the parameter $U_0$ is irrelevant at zero temperature) by minimising $H$ given in Eq.~\eqref{gibbs2} with respect to the lattice parameters $\phi$, $\gamma$, $\ell$, $\alpha$ and $\beta$. Since this is a relatively high dimensional minimisation, the minimisation proceeded via several stages. We first minimised $H$ over a relatively wide range of values for the lattice parameters to obtain an initial estimate for their equilibrium values. We then further minimised $H$ over a much smaller range around these initial estimates to obtain more refined estimates for the equilibrium lattice parameters. Finally, these refined estimates were used to guide us in finding exact values for the equilibrium lattice parameters using geometry (see Supplementary Information).

\subsection{Monte Carlo Simulations}\label{MC}

Monte Carlo (MC) Simulations were performed in the NVT ensemble in order to mimic the experiments in ref.\cite{Rey2017,Rey2018} where monolayer area rather than surface pressure was controlled. The simulations were carried out using a standard Metropolis scheme for $1024$ particles contained in a rectangular box with aspect ratio $2:\sqrt{3}$ and periodic boundary conditions. One of the challenges of studying the thin shell regime is the fact that the area fraction of the hard-cores is high, making it difficult to equilibrate the system.\cite{Jagla1998} In order to overcome this problem, long simulations and very gradual quenches were used. Specifically, for each particle area fraction, the particles were first disordered at $T^*=\infty$ (i.e., hard core repulsion only) and then brought to the final reduced temperature of $T^* = 0.01$ through successive stages of $T^*=0.3, 0.2, 0.1, 0.06, 0.03, 0.01$. At each stage, the system was equilibrated for $10^5$ attempted moves per particle with an acceptance probability of around $30\%$.

\section{Results \& Discussions}

\subsection{Phase Behaviour in the $g-P^*$ Plane}

Figure 1(c) shows the zero temperature phase diagram in the $g$ versus reduced pressure $P^*=r_0^2 P/U_0$ plane for $r_1/r_0 = 1.5$ while Figure 1(d) shows the corresponding MEC structures.

For $g \geq 1$, the soft shoulder profile is flat enough so that (under compression) it is energetically favourable for neighbouring shells to be either fully overlapped or not overlapped. This is shown for example by the fact that for $g \geq 1$, the hexagonal phase at low pressures corresponds to the low density hexagonal phase (HEXL, no overlap of corona) and at high pressures to the high density hexagonal phase (HEXH, full overlap of corona). This simple interplay between the hard-core and soft-shoulder length scales also leads to anisotropic MECs at intermediate pressures where all the lattice parameters of the MEC can be calculated exactly from geometry for $g \geq 1$ (see Supplementary Information).

For $g \geq 2$, the phase boundaries in Figure 1(c) become independent of $g$, with the same sequence of phases being observed as we increase $P^*$, going from the low density hexagonal phase (HEXL) to chains (CH), to what we shall call the zig-zag 1 phase (ZZ1) and finally to high density hexagons (HEXH), see Figure 1(d). On the other hand, for $1 \leq g \leq 2$, an additional phase, which we shall call the zig-zag 2 phase (ZZ2), emerges in between HEXH and ZZ1. Both ZZ1 and ZZ2 contain two particles per unit cell (see Figure 1 in Supplementary Information) and have their particles aligned along chains. However, the chains in both phases are not straight like in the CH phase, but have a zig-zag shape in order to accommodate the larger number of shell overlaps arising from the higher compression. Alternatively, ZZ1 and ZZ2 can be viewed as being made up of elongated six-particle rings, and these six-particle rings are a characteristic motif for both zig-zag phases. The main difference between ZZ1 and ZZ2 is the fact that the chains are slightly staggered with respect to each other in ZZ1 (so that there is no rectangular order), while they are in register with each other in ZZ2 (so that there is rectangular order), see Figure 1(d).

In contrast, for $g<1$, the repulsive shoulder is concave enough so that partial overlap of the corona becomes energetically favourable. This leads to more subtle MECs where only some lattice parameters can be fixed by geometry, whilst others have to be obtained via numerical minimization of the enthalpy. Specifically, the region $0.6 <g < 1$ represents a transitional zone where rhomboidal (RH) and square (SQ) phases with non-touching cores become more energetically favourable than the phases listed in the Supplementary Information. Finally, anisotropic structures disappear altogether for $g < 0.6$ so that the hexagonal phase goes continuously from HEXL to HEXH as we increase compression.

The broad features of the phase diagram shown in Figure 1(c) agree with the corresponding phase diagram calculated by Jagla in ref.\cite{Jagla1999} though there are also some important differences. Firstly, while the square phase (SQ) in the Jagla phase diagram extends considerably into the $g>1$ region (up to $g \approx 1.5$), it is confined to the $g<1$ region in Figure 1(c). Secondly, the ZZ1 phase is absent from Jagla's phase diagram. Finally, Jagla predicts the existence of an intricate phase containing five particles per unit cell (i.e., the S4 phase in ref.\cite{Jagla1999}) for $g \geq 1.5$, which is obviously absent from our phase diagram as we only consider up to two particles per unit cell. However, we note that the presence of the ZZ1 phase and absence of the S4 phase are both confirmed by very accurate Genetic Algorithm (GA) calculations containing up to 15 particles per unit cell by Fornleitner and Kahl for $r_1/r_0 = 1.5$ and $g=\infty$,\cite{Fornleitner2008,Fornleitner2010} providing independent validation for the accuracy of our two particle per unit cell calculations for the above state point. We believe that the discrepancy between our phase diagram and Jagla's is due to the incomplete set of MECs used by Jagla.

To verify that the zero temperature phase diagram in Figure 1(c) is correct, we performed Monte Carlo simulations along the dotted line ($g=0.75$), dashed line ($g=1$) and dotted-dashed line ($g=2$) in Figure 1(c) for $r_1/r_0=1.5$ and reduced temperature of $T^*=0.01$. Representative structures at different core area fractions $\eta=\rho \pi r_0^2/4$ along each of these lines are shown in Figure 2, where $\rho$ is the number density of the HCSS particles. In the insets of Figure 2, we also show a magnified view of selected regions of each snapshot to highlight local structure and help identify the underlying symmetry of the structure.

Note that the phase diagram in Figure 1(c) is calculated in the NPT ensemble, where the phase behaviour is controlled by changing pressure $P^*$, while the MC simulations in Figure 2 are performed in the NVT ensemble, where phase behaviour is controlled by changing density $\eta$. However, it is straightforward to find the pressure $P^*$ corresponding to different densities $\eta$ and vice-versa. Specifically, the coexistence pressure between two phases (i.e., the pressure at the phase boundary between the two phases) corresponds to $\eta$ lying between the $\eta$ values for the two phases. For example, for $g=1$, $r_1/r_0=1.5$ (i.e., dashed line in Figure 1(c)), the coexistence pressure between HEXL and CH is $P^*_{coex}=1.871$, the coexistence pressure between CH and ZZ1 is $P^*_{coex}=2.369$, while $\eta_{HEXL}=0.403$, $\eta_{CH}=0.555$ and $\eta_{ZZ1}=0.653$ (see Table 1 in Supplementary Information). Therefore, the pressure of the system is $P^*=1.871$ in the density range $0.403 < \eta <0.555$ while $P^*=2.369$ in the density range $0.555 < \eta <0.653$. On the other hand, for $\eta=0.403$, the pressure of the system can be any value in the range $P^*<1.871$, while for $\eta=0.555$, the pressure of the system can be any value in the range $1.871 < P^* < 2.369$ etc..

For $g=0.75$, as we increase compression, the system transitions from HEXL to RH (unit cell angle $\phi \neq 90^\circ$) to SQ to HEXH, see Figures 2(a)-(d) respectively, noting in particular the local structure shown in the insets. Apart from the higher density RH phase which we could not identify from the MC simulations, this sequence of structures is exactly that predicted by Figure 1(c) along the dotted line.

\begin{figure*}
 \centering
 \includegraphics[width=1.8\columnwidth]{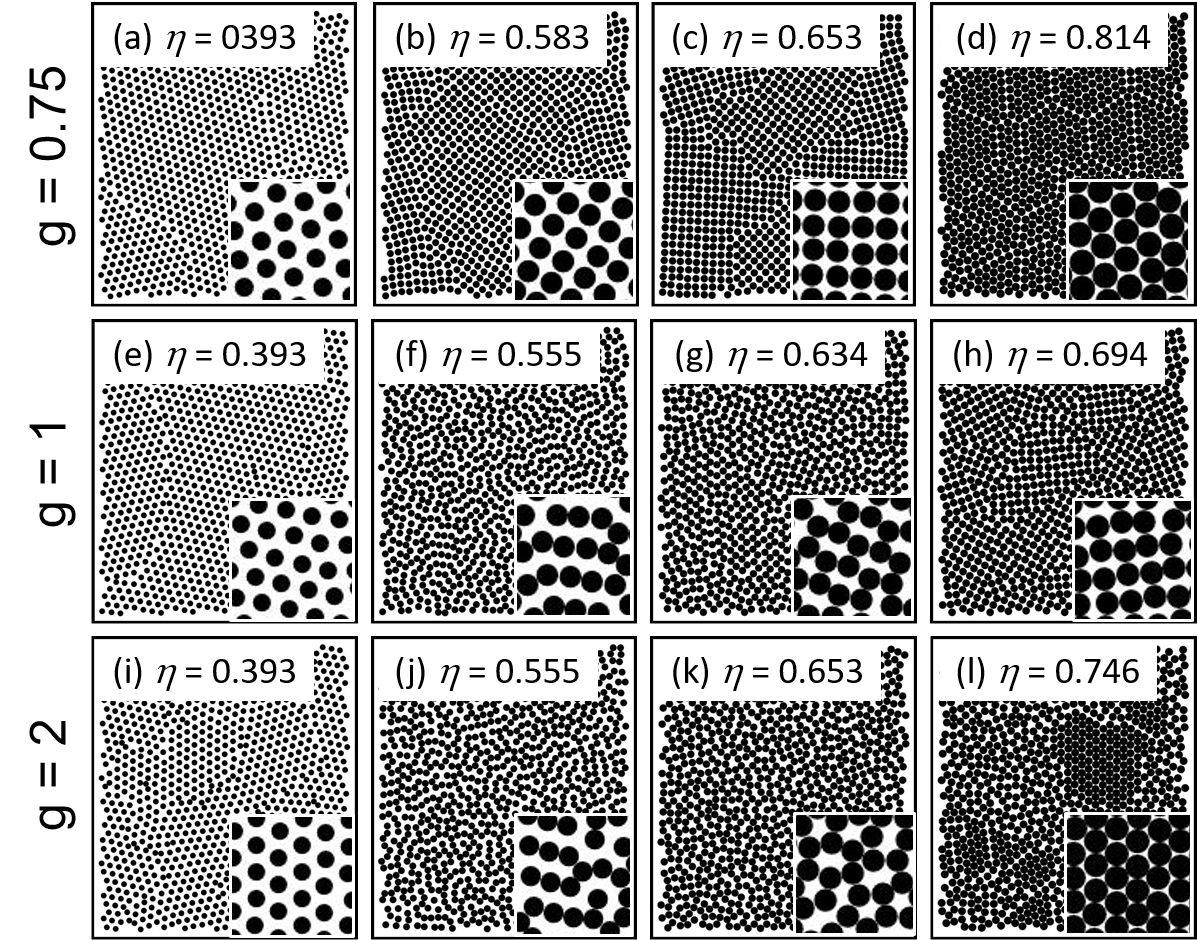}
 \caption{(a-l) Snapshots from Monte Carlo simulations of HCSS particles for $r_1/r_0=1.5$, reduced temperature of $T^*=0.01$ and different values of $g$ and core area fractions $\eta$. Note that the $g=0.75$, $g=1$ and $g=2$ simulations correspond to the dotted, dashed and dotted-dashed vertical lines respectively in Figure 1(c). See main text for a discussion of how to find the reduced pressure $P^*$ corresponding the different values of $\eta$ in the snapshots. The solid disks in the snapshots represent the particle cores while particle shells are omitted for clarity. In the insets, we show a magnified view of selected regions of each snapshot to highlight local structure.}
 \label{figure2}
\end{figure*}

For $g=1$, as we increase compression, the system transitions from HEXL to CH to ZZ1 to ZZ2, see Figures 2(e)-(h) respectively, and finally to a coexistence between ZZ2 and HEXH at the highest density we could access (not shown). We can identify that Figures 2(g) and (h) correspond to zig-zag phases (either ZZ1 or ZZ2) from the fact that the characteristic elongated six-particle ring motif is present in the local structure (see insets). We can further distinguish between ZZ1 and ZZ2 in these snapshots from the fact that there is no evident rectangular order in Figure 2(g) main panel (i.e., the particle chains are staggered, hence ZZ1), while the rectangular order is evident in Figure 2(h) main panel (i.e., the particle chains are in register, hence ZZ2). In Figure 2(f), the system is in the CH phase as can be seen from the local structure in the inset (compare this to the CH structure in Figure 1(d)). We note that the sequence of structures shown in Figure 2(e)-(h) is exactly that predicted by Figure 1(c) along the dotted line.

For $g=2$, as we increase compression, the system transitions from HEXL to CH to ZZ1 to a coexistence between ZZ1 and HEXH at the highest density we could access, see Figure 2(i)-(l) respectively. Note that the ZZ1 phase in Figures 2(k),(l) are much less ordered than in Figure 2(g) but the phase can still be discerned from isolated instances of the characteristic six-particle rings in the local structure (see for example the inset of Figure 2(k)). The sequence of structures shown in Figure 2(i)-(l) is exactly that predicted by Figure 1(c) along the dotted-dashed line.

Finally, we note that the core area fractions of all the snapshots in Figure 2 are in good agreement with the area fractions of the corresponding MECs given in Table 1 in the Supplementary Information. The MC results in Figure 2 therefore confirm the accuracy of the phase diagram in Figure 1(c).

\subsection{Phase Behaviour in the $r_1/r_0 - P^*$ Plane}

In Figures 3(a),(b), we show the zero temperature phase diagram in the $r_1/r_0$ vs reduced pressure $P^*$ plane for $g=1$ and $g=10$ respectively. These values of $g$ were chosen because $g=1$ is relevant to experimental two-dimensional core-shell particles\cite{Rey2017,Rey2018} while $g=10$ is essentially equivalent to the square shoulder repulsion $g=\infty$ (see Figure 1(a)) which has been extensively studied theoretically in the literature.

\begin{figure}[h]
\centering
  \includegraphics[width=1.0\columnwidth]{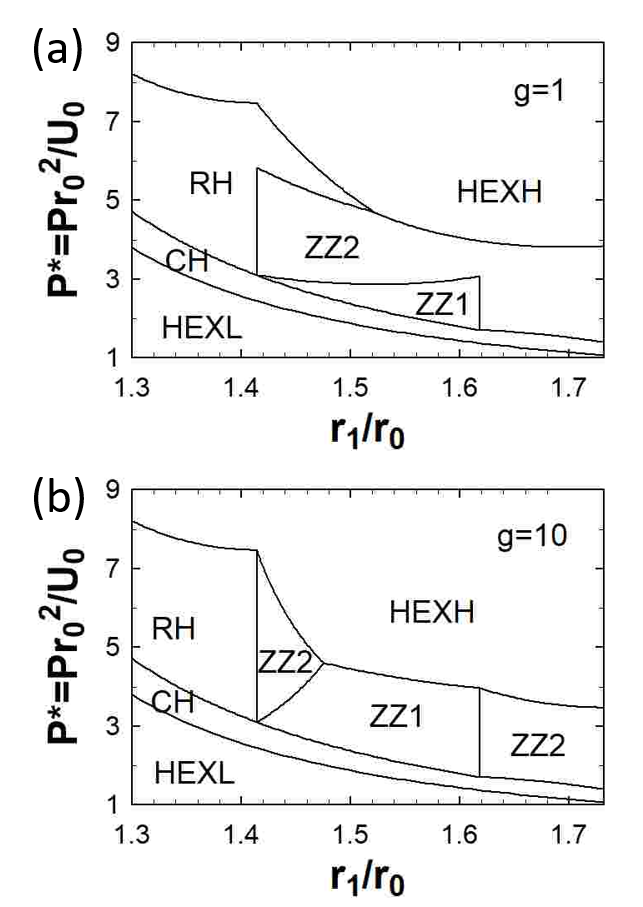}
  \caption{Zero temperature phase diagram for core-shell particles in the $r_1/r_0$--$P^*$ plane for (a) $g=1$ (b) $g=10$. The minimum energy configurations (MECs) labelled in the phase diagrams are low density hexagons (HEXL, no overlap of corona), high density hexagons (HEXH, full overlap of corona), chains (CH), rhomboids (RH), zig-zag 1 (ZZ1) and zig-zag 2 (ZZ2). See main text for more details.}
  \label{figure3}
\end{figure}

We see that there are some differences between the phase diagrams in Figures 3(a) and 3(b), e.g., for $g=\infty$ the ZZ1 phase occupies a larger area of the phase diagram (at the expense of the ZZ2 phase) compared to $g=1$. However, apart from these relatively minor differences, the two phase diagrams are broadly similar. In what follows, we therefore focus on the phase behaviour of HCSS particles with $g=1$ (linear ramp soft shoulder) since this potential is more realistic experimentally.\cite{Rey2017,Rey2018}

In Figures 3(a), several of the phase boundaries are vertical, occurring at certain significant values of $r_1/r_0$. These correspond to values of $r_1/r_0$ where changes in the geometry of the particle arrangements occur. For example $r_1/r_0=\sqrt{2}$ forms the phase boundary between RH and ZZ2. For this value of $r_1/r_0$, both the RH and ZZ2 phases approach the square structure so that there is a continuous phase transition from RH to ZZ2 as we increase $r_1/r_0$. The formation of the square lattices for HCSS particles with $g=1$, $r_1/r_0 \approx \sqrt{2}$ has been confirmed in our previous MC simulations.\cite{Rey2018}

More interestingly, $r_1/r_0 = \sqrt{2}$ corresponds to one of the special shell to core ratio where the HCSS particles can form quasicrystals.\cite{Dotera2014} This is because for $r_1/r_0=\sqrt{2}$, the unit cell of the RH phase is a square with side length $r_0$ while the unit cell of the HEXH phase consists of two compact equilateral triangles with side length $r_0$. At the phase boundary between the RH and HEXH phase (i.e., $P^*=7.464$ for $g=1$ and $r_1/r_2=\sqrt{2}$, see Figure 3(a)), the enthalpy per particle of both phases are equal. In the $NVT$ ensemble, the RH and HEXH phases are in coexistence for the core area fraction range $0.785 \leq \eta \leq 0.907$ (see Table 1 in Supplementary Information). Random tilings of squares and triangles can therefore occur in this range\cite{Henley1988} and the resultant random tiling phases are degenerate ground states of the system at $T^*=0$.\cite{Jagla1998} When the number ratio of squares to triangles is $\sqrt{3}/4$, the structure obtained is a random dodecagonal (12-fold symmetric) quasicrystal.\cite{Leung1989,Kawamura1991,Widom1993} Taking into account the fact that a square and an equilateral triangle contain $1$ particle and $\frac{1}{2}$ a particle respectively, we expect to find dodecagonal quasicrystals for core area fractions around $\eta \approx 0.848$ at $T^*=0$.

\begin{figure*}
 \centering
 \includegraphics[width=1.8\columnwidth]{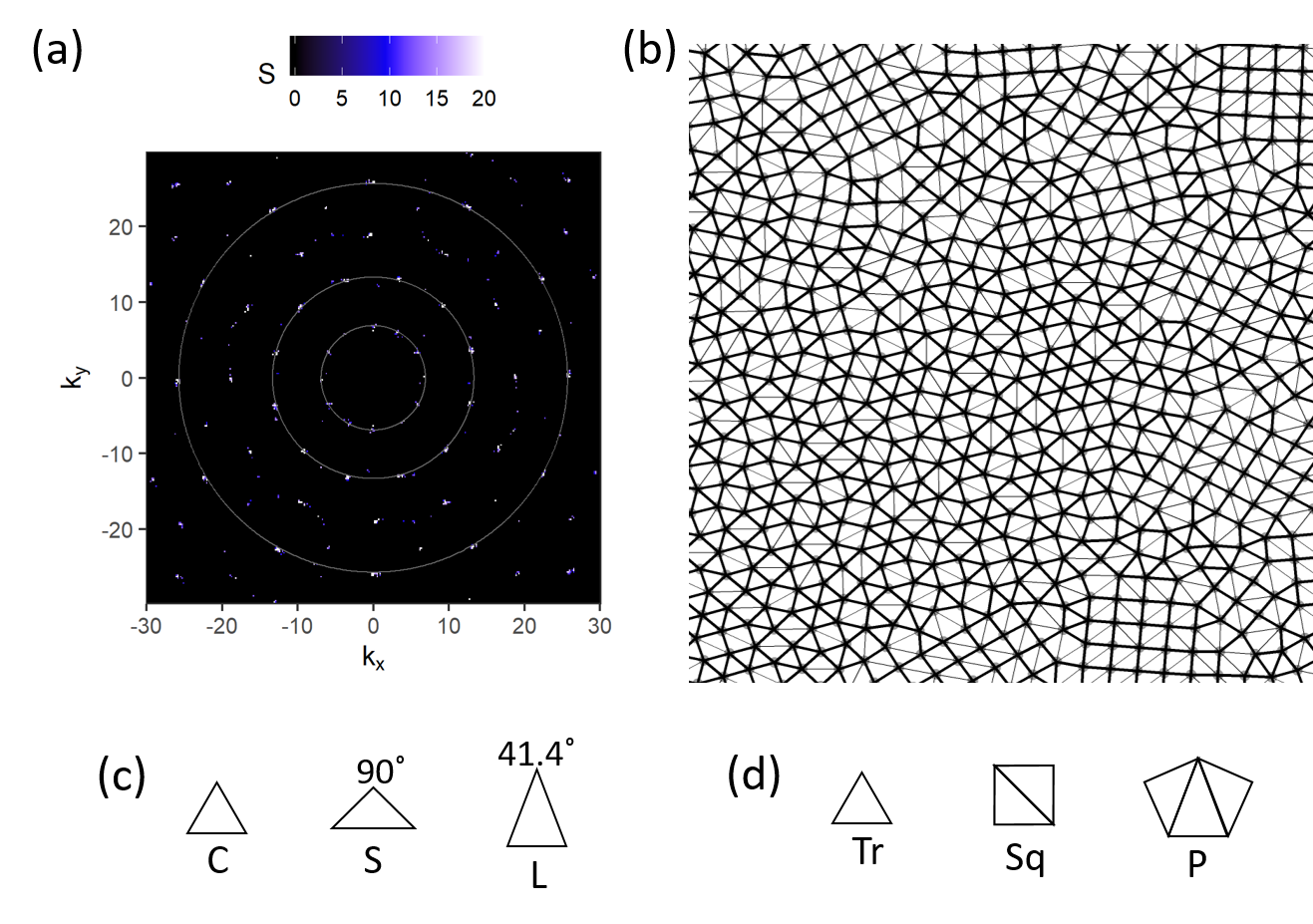}
 \caption{(a) Diffraction pattern calculated from Monte Carlo (MC) simulations of HCSS particles with $r_1/r_0=1.41$, $\eta=0.778$, $T^*=0.01$ and $g=1$. The innermost circle corresponds to the wavenumber where the first set of 12-fold diffraction peaks occurs while the radius ratio for successive circles is 1.93. (b) Real space image of a MC snapshot of the system where the particle centres are denoted by small grey disks while the thin lines are the Delaunay triangulation of the particle centres. We have also drawn thick lines between particles whose centres are closer than $1.3 r_0$ to accentuate the polygonal tiles in the system. (c) The Delaunay triangles found in (b), including compact equilateral triangles (C) and short (S) and long (L) isosceles triangles. (d) Some of the tiles found in (b), compact equilateral triangles (Tr), squares (Sq) and pentagons whose internal angles are not $108^\circ$ (P).}
 \label{figure4}
\end{figure*}

In Figure 4(a), we show the 2D diffraction pattern calculated from MC simulations of HCSS particles with $r_1/r_0=1.41$, $\eta=0.778$, $T^*=0.01$ and $g=1$. The diffraction pattern clearly shows 12-fold symmetry, indicating that the system is a dodecagonal quasicrystal. In addition, the wavenumber ratio for the successive circles of 12-fold peaks shown in Figure 4(a) is 1.93, further confirming the dodecagonal symmetry.\cite{Rucklidge2012,Barkan2014,Lifshitz1997} Our results agree with those of Dotera et al\cite{Dotera2014} who obtained 12-fold quasicrystals for HCSS particles with $r_1/r_0=1.4$, $\eta=0.77$, $T^*=0.278$ and $g=\infty$ (i.e., square shoulder repulsion). Interestingly, the core area fractions for dodecagonal quasicrystals in both studies are considerably lower than the $T^*=0$ value of $\eta \approx 0.848$. This fact is consistent with the study of Pattabhiraman et al \cite{Pattabhiraman2015} who found that for HCSS particles with $r_1/r_0=1.4$, configurational entropy causes the dodecagonal quasicrystal phase to shift to slightly lower core area fractions at finite temperatures.

In order to analyse the real-space structure of the quasicrystal, in Figure 4(b) we show the Delaunay triangulation of the particle centres from a MC snapshot of our system (thin lines between particle centres). We see that the Delaunay triangles primarily consist of compact equilateral triangles with side length $r_0$ (C), short isosceles triangles with apex angle $90^\circ$ and side length $r_0$ for the two equal sides (S) and long isosceles triangles with apex angle $41.4^\circ$ and side length $r_1$ for the two equal sides (L)(see Figure 4(c)). Note that two S triangles can be combined along their long edge to form a square (See Figure 4(d)).

Another instructive way to analyse the real space structure is to break the pattern up into tiles by joining together particle pairs whose cores are close to contact. Specifically, in Figure 4(b) we join together particle pairs whose centres are closer than $1.3 r_0$ (thick lines). We see that the resultant tiles consist mainly of compact equilateral triangles (Tr) and squares of side length $r_0$ (Sq), together with a small number of pentagons whose internal angles are not $108^\circ$ (P) and irregular polygons,\cite{Dotera2014} see Figure 4(d). The predominance of square and triangular tiles is not surprising since, as discussed earlier, squares and equilateral triangles can be used as the fundamental building blocks for constructing random dodecagonal quasicrystals.\cite{Dotera2014,Leung1989,Widom1993,Oxborrow1993}

Our results in Figure 4 demonstrate that HCSS particles with a linear ramp shoulder repulsion can form 12-fold quasicrystals. These results are significant since in our previous study,\cite{Rey2017} we showed that the phase behaviour of mixed monolayers of polystyrene microspheres (diameter 1.5 $\mu$m) and poly(N-isoprpylacrylamide) (PNiPAm) microgels (diameter 150 nm) at the air/water interface could be quantitatively modelled using HCSS particles with $g=1$ and $r_1/r_0 \approx 1.4$. Our study suggests that when compressed to a suitable density, this system could form 12-fold quasicrystals. One important caveat here is the fact that the experimental system in ref.\cite{Rey2017} is not ergodic because the energy scale of the soft shoulder repulsion is much greater than $k_BT$ and the density of the 12-fold quasicrystalline state is relatively high. It may therefore be necessary to provide external energy input into the system, e.g., through mechanical vibrations, in order to help the system to find its equilibrium state.

Another significant value for $r_1/r_0$ is the golden ratio $r_1/r_0=\tau \equiv (1+\sqrt{5})/2 \approx 1.618$ which forms the phase boundary between ZZ1 and ZZ2. For this core-shell ratio, both ZZ1 and ZZ2 approach the same structure (see Figure 5(b)) so that there is a continuous phase transition from ZZ1 to ZZ2 as we increase $r_1/r_0$. The reason why the ZZ1 phase comes abruptly to an end at this point is because for $r_1/r_0>1.618$, the lattice constant $b$ in Figure 1(e) in Supplementary Information becomes smaller than $r_1$, leading to an additional overlap of two shells which is energetically unfavourable.

\begin{figure}[h]
\centering
  \includegraphics[width=1.0\columnwidth]{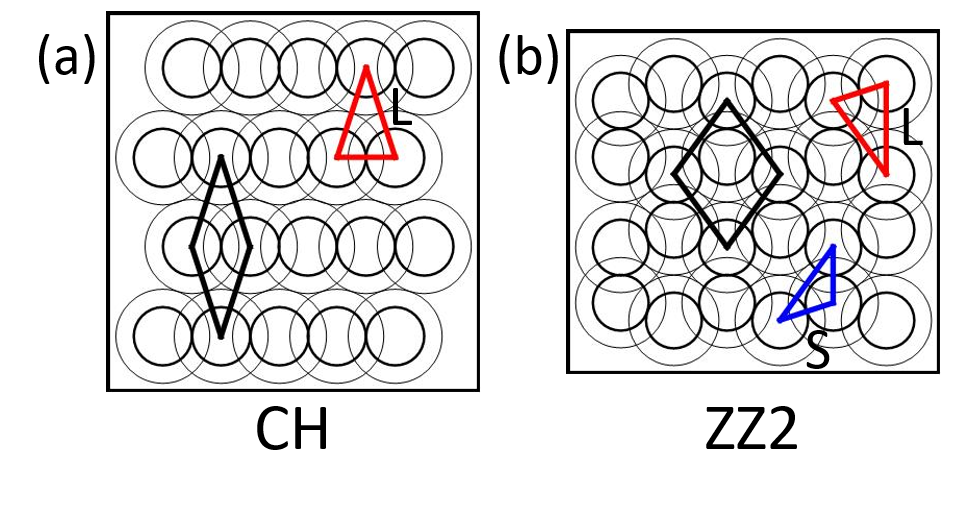}
  \caption{(a) CH phase for $r_1/r_0=1.618$, which can be built up from thin Penrose tiles (black rhombi), or more fundamentally from long Robinson triangles L (red triangle) shown in Figure 6(c). Note that the thin Penrose tile can be built up from two L triangles. (b) ZZ2 phase for $r_1/r_0=1.618$, which can be built up from fat Penrose tiles (black rhombi), or more fundamentally from both long (L) and short (S) Robinson triangles (red and blue triangles respectively) shown in Figure 6(c). Note the fat Penrose tile can be built up from two L and two S triangles.}
  \label{figure5}
\end{figure}

More interestingly, $r_1/r_0=\tau=1.618$ is another shell to core ratios where the system can form quasicrystals.\cite{Jagla1998,Dotera2014} This is because for $r_1/r_0=1.618$, we can use the fat and thin Penrose tiles as the unit cells for the ZZ2 phase and the CH phase respectively,\cite{Jagla1998} see black rhombi in Figure 5. Furthermore, at the phase boundary between ZZ2 and CH (i.e., $P^*=1.701$ for $g=1$, $r_1/r_2=1.618$, see Figure 3(a)), the enthalpy per particle of both phases are equal. In the $NVT$ ensemble, the ZZ2 and CH phases are in coexistence for the core area fraction range $0.510 \leq \eta \leq 0.631$ (see Table 1 in Supplementary Information), we therefore expect a random tiling of the two Penrose tiles to occur in this range. When the number fractions of fat and thin tiles are $\tau^{-1}$ and $\tau^{-2}$ respectively, the structure obtained is a random quasicrystal with decagonal (i.e., 10-fold) symmetry.\cite{Henley1988} Taking into account the fact that the fat and thin tiles in Figure 5 contain 2 particles and 1 particle respectively, this means that we should expect to find decagonal quasicrystals for core area fractions around $\eta \approx 0.60$.

In Figure 6(a), we show the 2D diffraction pattern calculated from MC simulation snapshots of HCSS particles with $r_1/r_0=1.618$, $\eta=0.59$, $T^*=0.01$ and $g=1$ (linear ramp potential). Notwithstanding the small degree of noise due to some disorder in the MC snapshot, the diffraction pattern clearly shows non-crystallographic 10-fold symmetry, indicating that the system is a decagonal quasicrystal. In addition, the wavenumber ratio for the successive circles of 10-fold peaks shown in Figure 6(a) is 1.618, further confirming the decagonal symmetry.\cite{Rucklidge2012,Barkan2014,Lifshitz1997} We note that 10-fold quasicrystals were also obtained by Jagla\cite{Jagla1998} for $r_1/r_0=1.65$, $P^*1.7$, $T^*<0.05$, $g=1$ and by Dotera et al\cite{Dotera2014} for $r_1/r_0=1.6$, $\eta=0.55$, $T^*=0.133$, $g=\infty$.

\begin{figure*}
 \centering
 \includegraphics[width=1.8\columnwidth]{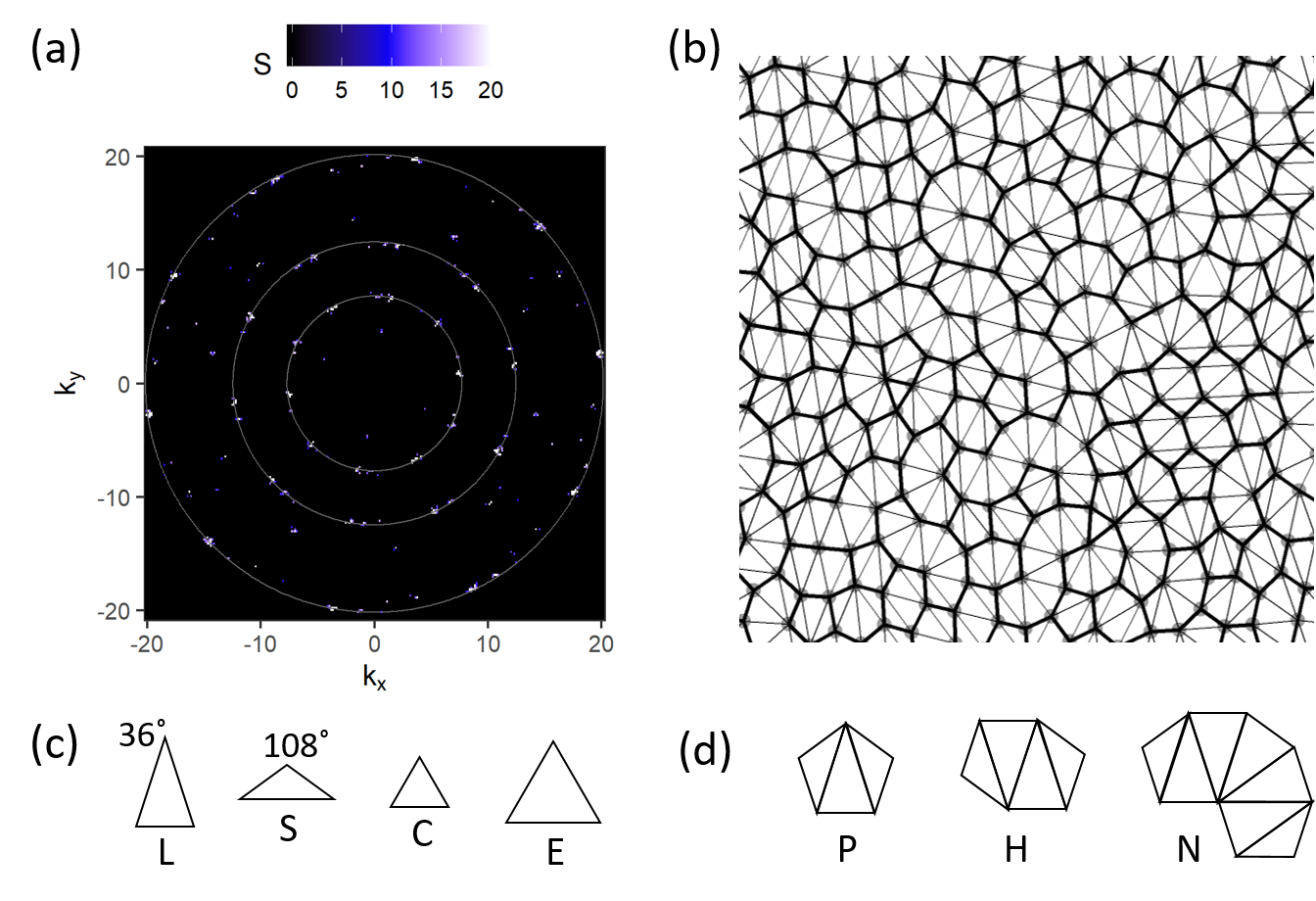}
 \caption{(a) Diffraction pattern calculated from Monte Carlo (MC) simulations of HCSS particles with $r_1/r_0=1.618$, $\eta=0.59$, $T^*=0.01$ and $g=1$. The innermost circle corresponds to the wavenumber where the first set of 10-fold diffraction peaks occurs while the radius ratio for successive circles is 1.618. (b) Real space image of a MC snapshot of the system where the particle centres are denoted by small grey disks while the thin black lines are the Delaunay triangulation of the particle centres. We have also drawn thick lines between particles whose centres are closer than $1.4 r_0$ to accentuate the polygonal tiles in the system. (c) The Delaunay triangles found in (b), including short (S) and long (L) Robinson triangles and compact (C) and expanded (E) equilateral triangles. (d) The standard decagonal tiles found in (b), including Pentagons (P), Hexagons (H) and Nonagons (N).}
 \label{figure6}
\end{figure*}

In order to analyse the real-space structure of the quasicrystal, in Figure 6(b) we show the Delaunay triangulation of the particle centres from a MC snapshot of our system (thin lines between particle centres). It is striking that the Delaunay triangles consist overwhelmingly of short (S) and long (L) Robinson triangles with apex angles of $108^\circ$ and $36^\circ$ respectively (see Figure 6(c)), and that these triangles also appear in both the CH and ZZ2 phases in Figure 5 (red (L) and blue (S) triangles). This is not surprising since Robinson triangles can be seen as the fundamental building blocks for constructing decagonal quasicrystals.\cite{Dotera2014} Indeed the thin Penrose tile in Figure 5(a) can be built up from two L triangles while the fat Penrose tile in Figure 5(b) can be built up from two L and two S triangles. In addition to the S and L triangles, the Delaunay triangles also consist of a very small number of compact equilateral triangles with side length $r_0$ (C) and expanded equilateral triangles with side length $r_1$ (E) (see Figure 6(c)).

As before, we also analyse the real space structure by joining together particle pairs whose centres are closer than $1.4 r_0$ (thick lines). We see that the resultant tiles are not in fact the idealised fat and thin Penrose tiles shown in Figure 5, but consist of a larger palette of standard decagonal tiles (including Pentagons (P), Hexagons (H), Nonagons (N), see Figure 6(d)),\cite{Dotera2014,Engel2007} less regular decagonal tiles and a small number of compact equilateral triangles (C). Apart from C, all the other tiles can essentially be built up from obtuse and acute Robinson triangles.

It is useful at this point to compare our results for the quasicrystalline phases with those of Schoberth et al\cite{Schoberth2016} which were also obtained for HCSS particles with non-square soft shoulder interactions. Specifically, these authors introduce a parameter $\alpha$ which controls the soft shoulder profile, with $\alpha = 0$ resembling the square shoulder potential and $\alpha \geq 10$ resembling the linear ramp potential. Interestingly, for linear-ramp-like potentials ($\alpha \geq 10$), these authors did not observe any quasicrystalline phases for shell to core ratios of $\lambda \approx 1.4$, while they observed an interesting mixed quasicrystal state (with relatively weak 10-fold symmetry locally and much stronger 12-fold symmetry on larger length scales) for $\lambda \approx 1.6$. These results are significantly different to our linear ramp results where we observe 12-fold quasi-crystals for $r_1/r_0 = 1.41$ and pure 10-fold quasi-crystals for $r_1/r_0 = 1.618$. The difference between the two sets of results suggest that the formation of quasicrystalline states may be relatively sensitive to subtle differences in the soft shell profile.

\begin{figure}[h]
\centering
  \includegraphics[width=0.8\columnwidth]{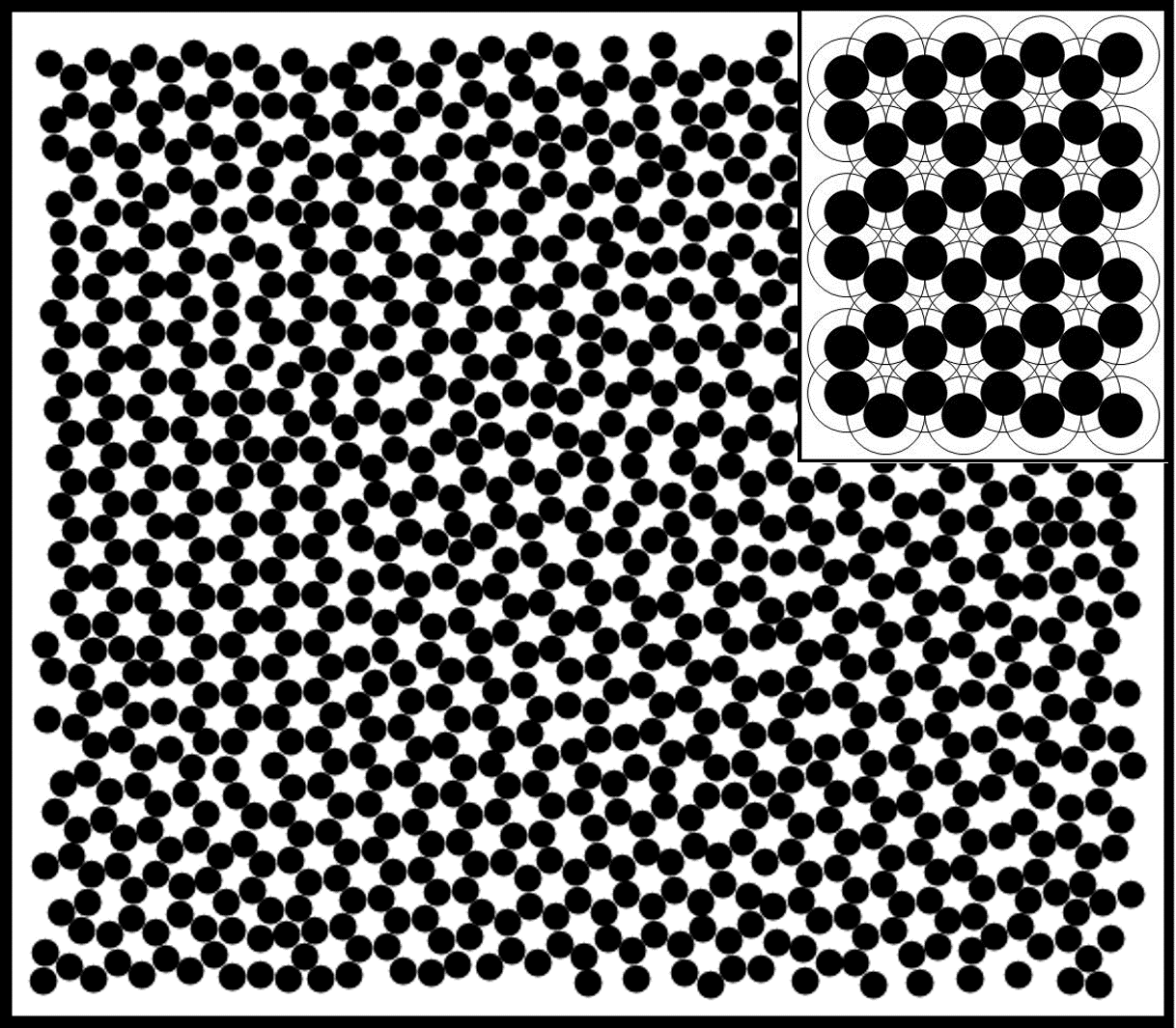}
  \caption{Snapshot from Monte Carlo simulations of HCSS particles with $r_1/r_0=1.73$, $g=1$, $\eta=0.605$ and $T^*=0.01$. Note the clear honeycomb structure in the snapshot. The inset shows the MEC for $r_1/r_0=1.73$.}
  \label{figure7}
\end{figure}

Finally, as $r_1/r_0 \rightarrow \sqrt{3}$, the ZZ2 structure approaches the honeycomb structure, see inset of Figure 7. To check whether this structure is accessible via MC simulations, in Figure 7, we show a representative snapshot from MC simulations of HCSS particles for $g=1$, $r_1/r_0=1.73$, $T^*=0.01$ and $\eta=0.605$ (the core area fraction of the ZZ2 phase for $r_1/r_0=1.73$, see Table 1). We clearly see the formation of honeycomb structures over large domains in agreement with the minimum energy calculations. Interestingly, honeycomb structures were also observed by Pattabhiraman and Dijkstra\cite{Pattabhiraman2017} for HCSS particles with a slightly larger shell to core ratio of $r_1/r_0=1.95$. However, the physics behind honeycomb formation is slightly different here as we are in the thin shell regime ($r_1/r_0 < \sqrt{3}$) where there are no shell overlaps beyond nearest neighbour particles.

\section{Conclusions}

We have studied the self-assembly of hard-core/soft shell (HCSS) particles in two dimensions and in the thin shell regime using both minimum energy calculations and Monte Carlo simulations. In contrast to most studies in this area, we have considered a range soft shell potential profiles beyond the standard square shoulder potential.

Our results for the phase diagram of HCSS particles agree with the broad features found in the seminal studies by Jagla,\cite{Jagla1998,Jagla1999} but also correct a number of important discrepancies in those studies. In particular, we show that by tuning the soft shoulder profile, the shell to core ratio $r_1/r_0$ and density, it is possible to generate hexagonal close packed structures, chains, rhomboids, squares and two distinct zig-zag structures. We also show that HCSS particles in the thin shell regime can form honeycombs when $r_1/r_0 \rightarrow \sqrt{3}$.

In addition to periodic structures, we show that by tuning both shell to core ratios and densities, HCSS particles with a linear ramp soft shoulder potentials can be engineered to form a variety of quasicrystalline structures, including 10-fold quasicrystals for $r_1/r_0=1.618$ and 12-fold quasicrystals for $r_1/r_0=1.41$. These results agree with the previous results of Dotera et al for HCSS particles with a square shoulder repulsion.\cite{Dotera2014} However, since linear ramp potentials are more experimentally realistic,\cite{Rey2017} these results serve as an important road map for the experimental realization of complex assembly phases using isotropic building blocks.

\section*{Conflicts of interest}
There are no conflicts to declare.

\section*{Acknowledgements}
WRCS and DMAB acknowledge funding from the EPSRC (Grant number EP/L025078/1) and the European Union's Horizon 2020 research and innovation programme under grant agreement No 861950, project POSEIDON. They also acknowledge the Viper High Performance Computing facility of the University of Hull and its support team. MR and NV acknowledge funding from the Deutsche Forschungsgemeinschaft (DFG) (grant number VO 1824/6-1) and and the European Union's Horizon 2020 research and innovation programme under grant agreement No 861950, project POSEIDON. AJA acknowledges funding from the EPSRC (Grant number EP/P015689/1 Quasicrystals: how and why do they form?).

%%%END OF MAIN TEXT%%%

%The \balance command can be used to balance the columns on the final page if desired. It should be placed anywhere within the first column of the last page.

\balance

%If notes are included in your references you can change the title from 'References' to 'Notes and references' using the following command:
%\renewcommand\refname{Notes and references}

%%%REFERENCES%%%
%\bibliography{core-shell} %You need to replace "rsc" on this line with the name of your .bib file
%\bibliographystyle{rsc} %the RSC's .bst file

\providecommand*{\mcitethebibliography}{\thebibliography}
\csname @ifundefined\endcsname{endmcitethebibliography}
{\let\endmcitethebibliography\endthebibliography}{}

\end{document}